\pdfoutput=1
\documentclass[fleqn,usenatbib]{mnras}
\usepackage{graphicx}
\usepackage{amsmath,amssymb}
\usepackage{lastpage}

\usepackage[all]{hypcap}
\usepackage[T1]{fontenc}

\usepackage{float}
\usepackage{subfigure}
\usepackage{afterpage}
\usepackage[usenames]{color}
\usepackage{yfonts}
\usepackage{mathrsfs}
\usepackage{upgreek}
\usepackage{hyperref}

\newcommand{\red}{\textcolor{black}}

\title{The scatter, residual correlations and curvature of the \textsc{sparc} baryonic Tully--Fisher relation}
\author[H.~Desmond]{
Harry~Desmond$^{1,2}$\thanks{E-mail: harryd2@stanford.edu}
\\
$^{1}$Kavli Institute for Particle Astrophysics and Cosmology, Physics Department, Stanford University, Stanford, CA 94305, USA\\
$^{2}$SLAC National Accelerator Laboratory, Menlo Park, CA 94025, USA\\
}

\pubyear{2017}

\begin{document}
\label{FirstPage}
\pagerange{\pageref{FirstPage}--\pageref{LastPage}}
\maketitle

\begin{abstract}
In recent work,~\citet{Lelli} argue that the tightness of the baryonic Tully--Fisher relation (BTFR) of the \textsc{sparc} galaxy sample, and the weakness of the correlation of its residuals with effective radius, pose challenges to $\Lambda$CDM cosmology. In this \textit{Letter} we calculate the statistical significance of these results in the framework of halo abundance matching, which imposes a canonical galaxy--halo connection. Taking full account of sample variance among \textsc{sparc}-like realisations of the parent halo population, we find the scatter in the predicted BTFR to be $3.6\:\sigma$ too high, but the correlation of its residuals with galaxy size to be naturally weak. Further, we find abundance matching to generate BTFR curvature in $3.0\:\sigma$ disagreement with the data, and a fraction of galaxies with non-flat rotation curves somewhat larger than observed.
\end{abstract}


\section{Introduction}
\label{sec:intro}

Among galaxy scalings, the correlation of baryonic mass with rotation velocity (baryonic Tully--Fisher relation; BTFR) stands out. In addition to possessing a very small intrinsic scatter, the BTFR is an almost perfect power-law over six decades of mass, describes galaxies with a wide range of morphologies, and has residuals systematically uncorrelated with other galaxy variables. This makes it at once a strong test of galaxy formation theories and an important source of information on their degrees of freedom.

Recently,~\citet[hereafter L16]{Lelli} have presented the BTFR of the \textsc{sparc} sample~\citep{SPARC}, a compilation of $175$ galaxies with high-quality H\textsc{i} rotation curves (RCs) and \textit{Spitzer} imaging at $3.6 \, \mu m$. The authors claim two features of the \textsc{sparc} BTFR to be very difficult for $\Lambda$CDM-based models to account for: its small intrinsic scatter $s_\text{BTFR}$ ($\sim0.11$ dex in baryonic mass) and the negligible correlation $\rho$ of its residuals with galaxy size. In standard galaxy formation, $s_\text{BTFR}$ should receive contributions from the halo mass--concentration and halo mass--galaxy mass relations, both of which themselves have scatter $0.1-0.2$ dex, and a simple application of Kepler's laws may be expected to yield an anticorrelation of velocity and size residuals.

Although valid, these arguments lack the statistical evidence required to claim a significant discrepancy. The aim of this work is to supply that evidence. In particular, we will calculate the expectation for $s_\text{BTFR}$ and $\rho$ in a vanilla $\Lambda$CDM model described by abundance matching (AM) by constructing mock data sets identical to \textsc{sparc} in all baryonic variables and analysed in precisely the same way. We will find two additional effects to be important, neither of which have previously been considered in detail: 1) sample variance in BTFR statistics between \textsc{sparc}-like realisations of the full galaxy--halo population, and 2) the falloff with galactocentric radius of the sensitivity of the baryonic component of the RC to galaxy size. We will show that when these effects are accounted for in a complete and fully self-consistent comparison with the \textsc{sparc} data, the discrepancies in $s_\text{BTFR}$ and $\rho$ are $\sim3.6\:\sigma$, and $2.2\:\sigma$ respectively. In addition, we investigate two further statistics that are constraining for galaxy formation models: the BTFR curvature and the fraction of RCs that are flat. The significance levels at which the \textsc{sparc} values for these quantities differ from those of the model are $\sim3.0\: \sigma$ and $2.2\: \sigma$. We conclude that the BTFR statistics of the \textsc{sparc} data pose a challenge to AM models that is moderately statistically significant.

\red{Our work builds on a number of studies aimed at assessing the consistency between the observed BTFR and the $\Lambda$CDM prediction, which have been carried out within both AM (e.g.~\citealt{TG, D12, DC}) and hydrodynamical (e.g.~\citealt{Santos-Santos, Dutton17}) frameworks. Despite a broad consensus that the general shape of the relation is compatible with $\Lambda$CDM, the precise extent of this agreement -- as well as the significance of more detailed features such as intrinsic scatter, curvature and residual correlations -- remains unclear. We intend our focused work on \textsc{sparc} to pave the way for more general statistical analyses in the future.}

\section{Method}
\label{sec:method}

Before detailing our procedure, we describe the twofold novelty of our approach. First, by using mock galaxies with baryonic properties identical to those of \textsc{sparc} and sampled at the same radii in the same ways, we eliminate systematic error in the comparison of BTFR statistics and ensure that any differences with the observed dynamics are due solely to the distribution of dark matter. Second, by thoroughly sampling the set of halo properties that may be associated with a given galaxy by AM, we robustly calculate the sample variance of each BTFR statistic in the model. Simple frequentist methods will then allow us to determine the significance of differences with the corresponding statistics in the data. Our method is similar to that of~\citet{D17}, on which it is based. The steps are the following.

\begin{enumerate}

\item{} From the full \textsc{sparc} data set, remove starburst dwarfs and galaxies with $i<30^{\circ}$ or quality flag 3. These are all the selection criteria of L16, except for a cut on RC flatness which we will come to shortly. We denote the resulting sample, containing $150$ galaxies, as ``\textsc{sparc}'' hereafter.

\item{} Estimate the true stellar and gas masses of each \textsc{sparc} galaxy by scattering the measured values (assuming $M_*/L=0.5$ for the disc, $M_*/L=0.7$ for the bulge, and $M_\text{gas}=1.33\:M_\text{H\textsc{i}}$;~\citealt{SPARC}) by the measurement uncertainties calculated using L16, eq. 5. Use the stellar mass to assign each galaxy a halo by the technique of abundance matching (AM;~\citealt{Kravtsov,Conroy}). In particular, we will use the AM model that~\citet{Lehmann} find to reproduce best the correlation function of SDSS, and match to halos in the \textsc{darksky-400} simulation~\citep{DarkSky}, a $(400 \: \mathrm{Mpc~h^{-1}})^3$ box with $4096^3$ particles run with the \textsc{2hot} code~\citep{Warren}. We identify halos using \textsc{rockstar}~\citep*{Rockstar}.

\item{} Use an NFW profile with the concentration and mass \red{(subtracting the baryon fraction)} of the assigned N-body halo to calculate the velocity due to the dark matter at each of the radii at which the RC of each \textsc{sparc} galaxy was probed~\citep{SPARC}. Add in quadrature the fixed baryonic contribution (imported directly from the \textsc{sparc} data) to calculate the total velocity, then scatter by the corresponding uncertainty (L16, eq. 3) to model observational error.

\item{} Use the algorithm of L16 (eqs. 1-2) to determine whether a given model galaxy has a flat RC, and if so to calculate the corresponding velocity $V_\text{f}$. If the RC is not flat, discard the galaxy. Denote by $N_\text{f}$ the total number of galaxies in the mock data set removed in this way.

\item{} Use $V_\text{f}$ and the baryonic masses ($M_\text{b}$) of the remaining galaxies to calculate the BTFR statistics. Begin by fitting to the BTFR and $M_\text{b}-R_\text{eff}$ relation quadratic curves in log-log space, with Gaussian scatter in $\log(V_\text{f})$ and $\log(R_\text{eff})$ respectively, by maximising the corresponding likelihood model. Subtract the $V_\text{f}$ and $M_\text{b}$ measurement uncertainties in quadrature from the total scatter to estimate the intrinsic scatter $s_\text{BTFR}$. Take the best-fitting coefficient of the quadratic term, $q$, as a measure of the BTFR curvature.

\item{} Calculate the velocity and radius residuals as

\begin{equation}
\Delta V_\text{f} \equiv V_\text{f} - \langle V_\text{f}|M_\text{b} \rangle
\end{equation}

\noindent and 

\begin{equation}
\Delta R_\text{eff} \equiv R_\text{eff} - \langle R_\text{eff}|M_\text{b} \rangle,
\end{equation}

\noindent where $\langle Y|M_\text{b} \rangle$ denotes the expectation for $Y$ at fixed $M_\text{b}$ given the fit to the full relation, and calculate $\rho$ as the Spearman's rank coefficient of their correlation. This completes the treatment of a single mock data set.

\item{} Repeat steps (ii)-(vi) for 2000 mock data sets, in each case randomly drawing for each \textsc{sparc} galaxy a different \textsc{darksky-400} halo consistent with the AM model. This generates distributions of $s_\text{BTFR}$, $\rho$, $N_\text{f}$ and $q$ that fully capture the sample variance of the model predictions.

\item{} Calculate the significance of the difference between model and data for each of $X \equiv \{s_\text{BTFR}, \: \rho, \: N_f, \: q\}$ as

\begin{equation}
\sigma_X \equiv (\langle X \rangle - X_d)/s_X,
\end{equation}

\noindent where $\langle X \rangle$ is the mean of $X$ over all mock data sets, $s_X$ is the standard deviation of the distribution, and $X_d$ is the corresponding value in the \textsc{sparc} data.

\end{enumerate}

\section{Results}
\label{sec:results}

We present the predicted vs observed \textsc{sparc} BTFR in Fig.~\ref{fig:btfr}, the correlation of $V_\text{f}$ and $R_\text{eff}$ residuals in Fig.~\ref{fig:res}, and the distribution of each statistic $X$ in Fig.~\ref{fig:hists}. Table~\ref{tab:table} lists the mean and standard deviations of these distributions, along with the significances of offsets from the data. Here we describe these results: Section~\ref{sec:residuals} focuses on $\rho$, Section~\ref{sec:scatter} on $s_\text{BTFR}$, Section~\ref{sec:N_f} on $N_\text{f}$ and Section~\ref{sec:curvature} on curvature $q$.

\subsection{The $\Delta{R_\textrm{eff}}-\Delta{V_\text{f}}$ correlation}
\label{sec:residuals}

\begin{figure*}
  \subfigure[]
  {
    \includegraphics[width=0.45\textwidth]{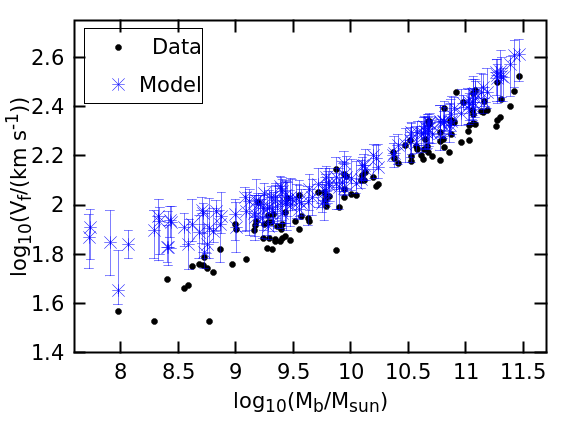}
    \label{fig:btfr}
  }
  \subfigure[]
  {
    \includegraphics[width=0.45\textwidth]{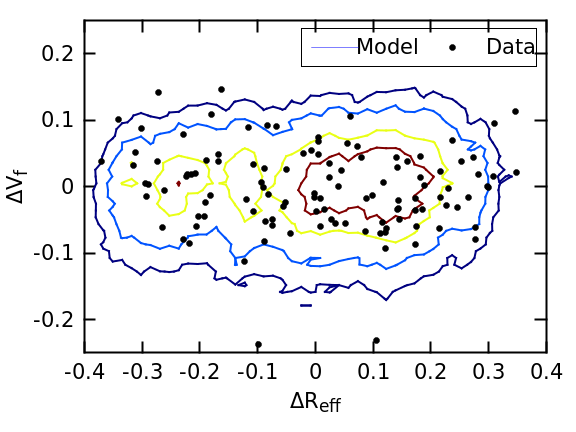}
    \label{fig:res}
  }
  \caption{The prediction of abundance matching applied to the \textsc{sparc} sample for the BTFR (Fig.~\ref{fig:btfr}) and $\Delta R_\text{eff}-\Delta V_\text{f}$ correlation (Fig.~\ref{fig:res}), compared to the data itself. Blue stars show the modal $V_\text{f}$ over all mock data sets for each \textsc{sparc} galaxy, and error bars show the $1\: \sigma$ variation. While the model BTFR is curved and has higher scatter than is observed, its residuals are correctly uncorrelated with galaxy size.}
  \label{fig:fig1}
\end{figure*}

\begin{figure*}
  \subfigure[]
  {
    \includegraphics[width=0.45\textwidth]{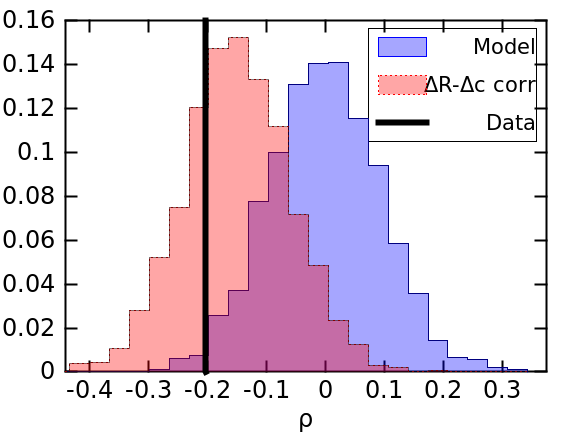}
    \label{fig:spear}
  }
  \subfigure[]
  {
    \includegraphics[width=0.45\textwidth]{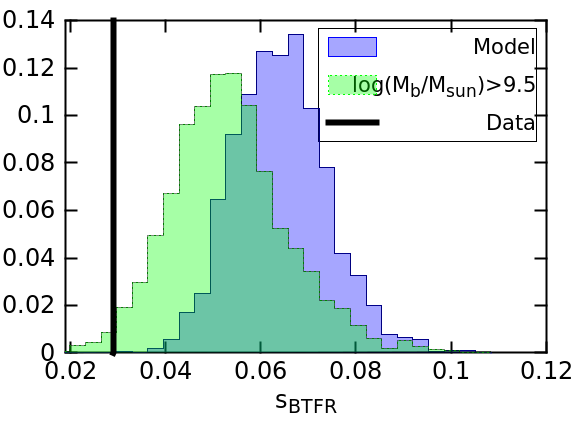}
    \label{fig:sig}
  }
  \subfigure[]
  {
    \includegraphics[width=0.45\textwidth]{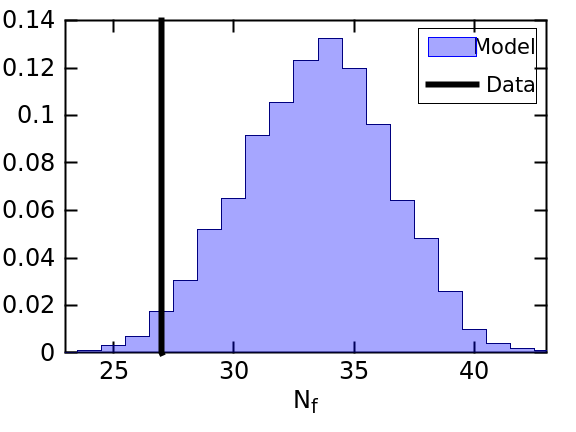}
    \label{fig:nfail}
  }
  \subfigure[]
  {
    \includegraphics[width=0.45\textwidth]{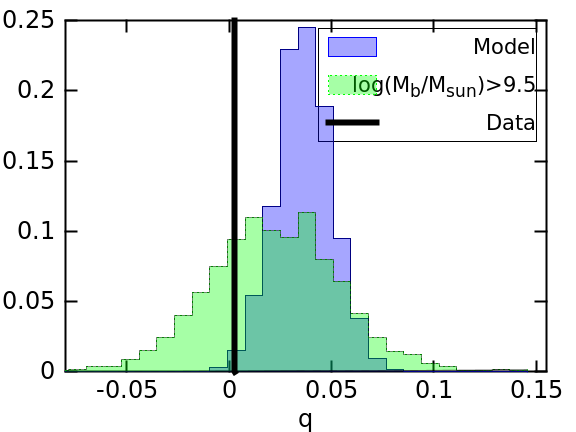}
    \label{fig:quad}
  }
  \caption{The distributions of four key BTFR statistics predicted by AM -- the strength of the $\Delta R_\text{eff}-\Delta V_\text{f}$ correlation $\rho$, the intrinsic scatter $s_\text{BTFR}$, the number $N_\text{f}$ of galaxies with non-flat RCs, and the curvature $q$ -- compared to the values in the \textsc{sparc} data. The results are quantified in Table~\ref{tab:table}. `$\Delta R$-$\Delta c$ corr' in Fig.~\ref{fig:spear} denotes an anticorrelation of $R_\text{eff}$ and $c$ residuals as described in Section~\ref{sec:residuals}.}
  \label{fig:hists}
\end{figure*}

We begin with the correlation of $R_\text{eff}$ and $V_\text{f}$ residuals. In Figure~\ref{fig:res} we stack $\Delta R_\text{eff}$ and $\Delta V_\text{f}$ of all mock data sets to form a contour plot, on which we overlay the \textsc{sparc} data. In Figure~\ref{fig:spear} we compare the distribution of $\rho$ in the mock data to the corresponding value in the real data, and in the first row of Table~\ref{tab:table} we report $\langle \rho \rangle$, $s_\rho$, $\rho_d$ and $\sigma_\rho$.

It is clear that the model prediction is not particularly discrepant with the data: neither show a strong $\Delta R_\text{eff}-\Delta V_\text{f}$ correlation. This may be understood as follows. $V_\text{f}$ is calculated from the flat part of the RC defined by the final measured points. As this is typically several times beyond $R_\text{eff}$ ($\sim2-10$ for the \textsc{sparc} sample), the velocity contribution due to the galaxy is effectively that of a point mass at its centre. Variations in galaxy size at fixed $M_\text{b}$ provide only a small perturbation to this leading order term, rendering $\langle \rho \rangle$ negligible. (Similar results obtain replacing $R_\text{eff}$ by the scale length of the stellar or gas disc.) Note that this is very different to the results of~\citet{DW15} (their figs. 6-7), which L16 cite as evidence for the expectation $\rho \ll 0$. This is because~\citet{DW15} use the velocity at the radius enclosing 80\% of the $i$-band light, where the baryonic contribution to the RC is not only larger but depends much more sensitively on $R_\text{eff}$.

In fact, $\rho_d$ is \emph{more} negative than $\langle \rho \rangle$, indicating a \emph{stronger} residual anticorrelation in the data than predicted by the model. Although only $2.2\:\sigma$ significant, this provides evidence within our framework for a second component of the galaxy--halo connection: an anticorrelation of $R_\text{eff}$ with halo mass $M_\text{vir}$ or concentration $c$ at fixed $M_\text{b}$. This would give smaller galaxies on average more dark matter within $R_\text{f}$, and hence larger $V_\text{f}$. Such a correlation has already been suggested by~\citet{D17} on the basis of the correlation of the residuals of the mass discrepancy--acceleration relation (MDAR) with galaxy size. The red histogram in Figure~\ref{fig:spear} shows the result for the best-fitting correlation found there, $\Delta R_\text{eff} \sim -0.4 \: \Delta c$; that this model gives a good fit to $\rho_d$ suggests the BTFR and MDAR to contain similar information in this regard. This correlation may however be in disagreement with the observed $\Delta R-\Delta V$ correlation when $V$ is measured further in~\citep{DW15}.

\begin{table}
  \begin{center}
    \begin{tabular}{l|r|r|r|}
      \hline
					Statistic & \textsc{sparc}	        & Model mean  		& Discrepancy/$\sigma$\\ 
      \hline
\rule{0pt}{3ex}
      $\rho$	& $-0.20$	& $0.00$         	& $2.2$\\
\rule{0pt}{3ex}
      $s_\text{BTFR}$ (dex)           	& $0.029$	& $0.064$        	& $3.6$\\
\rule{0pt}{3ex}
      \textquotedbl \: ($M_\text{b}>10^{9.5} M_\odot$)           	& $0.027$	& $0.053$        	& $2.1$\\
\rule{0pt}{3ex}
      $N_\text{f}$             		& $27$ 		& $33.4$         	& $2.2$\\
\rule{0pt}{3ex}
      $q$                 	& $0.003$	& $0.039$          	& $3.0$\\
      \hline
    \end{tabular}
  \caption{\red{Comparison of \textsc{sparc} and model BTFR statistics. $\rho$ is the Spearman's rank coefficient of the $\Delta R_\text{eff}-\Delta V_\text{f}$ correlation, $s_\text{BTFR}$ is the intrinsic BTFR scatter, $N_\text{f}$ is the number of galaxies with non-flat RCs, and $q$ is the quadratic BTFR curvature. The 3$^\text{rd}$ row shows $s_\text{BTFR}$ for $M_\text{b} > 10^{9.5} M_\odot$ galaxies only.}}
  \label{tab:table}
  \end{center}
\end{table}

\subsection{Scatter}
\label{sec:scatter}

We now proceed to $s_\text{BTFR}$. The blue histogram in Figure~\ref{fig:sig} shows the distribution of this statistic in the model, and the second row of Table~\ref{tab:table} lists $\langle s_\text{BTFR} \rangle$, $s_s$, $s_d$ and $\sigma_s$. As anticipated by L16\red{, and in agreement with~\citet{D12} and~\cite{DC}}, we find the predicted BTFR scatter to be upwards of 0.15 dex in $M_\text{b}$, with typical mock data sets having $\sim0.25$ dex.\footnote{The scatters in $M_\text{b}$ may be approximately obtained from the quoted scatters in $V_\text{f}$ by dividing by the BTFR slope, $\sim0.25$.} Given the spread among mock data sets, this is $3.6\: \sigma$ discrepant with the \textsc{sparc} value of $\sim0.11$ dex. This is significant -- none of our 2000 mock data sets have $s_\text{BTFR} < s_d$ -- but not phenomenally so. $s_\text{BTFR}$ can be reduced to a small degree by tightening the galaxy--halo connection: adopting an AM scatter of $0$ reduces $\langle s_\text{BTFR} \rangle$ to 0.061, with a corresponding discrepancy of $3.2\: \sigma$.

It is evident from Fig.~\ref{fig:btfr} that the predicted BTFR scatter rises towards lower $M_\text{b}$. To quantify this effect, we show in green in Figure~\ref{fig:sig} the $s_\text{BTFR}$ distribution with all $M_\text{b}<10^{9.5} M_\odot$ galaxies removed; this reduces the discrepancy to $2.1\: \sigma$ (Table~\ref{tab:table}, row 3). \red{The model prediction may be unreliable for $M_\text{b} \lesssim 10^{9.5} M_\odot$, as the stellar mass function requires extrapolation, AM cannot be directly tested with clustering, and low-mass halos may not be fully resolved}. We note also that gas mass fractions rise rapidly below $M_\text{b} \sim 10^{9.5} M_\odot$, amplifying any potential error incurred by performing AM with stellar as opposed to total baryonic mass. An $M_\text{b}$-based AM would correlate $M_\text{b}$ more strongly with $M_\text{vir}$ and $c$, likely reducing $s_\text{BTFR}$.

\subsection{Rotation curve flatness}
\label{sec:N_f}

Since we eliminate galaxies in each mock data set with non-flat RCs, Figs.~\ref{fig:fig1},~\ref{fig:spear},~\ref{fig:sig} and~\ref{fig:quad} pertain only to a subset of the full sample. An orthogonal statistic with which to compare model and data, therefore, is the number of galaxies out of the original $150$ that fail the flatness cut, which we denote $N_\text{f}$. Fig.~\ref{fig:nfail} shows the distribution of $N_\text{f}$ over all the mock data sets compared to the value in \textsc{sparc} (27), and the corresponding statistics are shown in the 4$^\text{th}$ row of Table~\ref{tab:table}. We find a larger fraction of our model galaxies to have rising RCs at the last measured point than in the data (a $2.2\: \sigma$ discrepancy), reflecting the fact the NFW density profile falls as $\sim r^{-1}$ out to large radius. This quantifies the longstanding  ``disc--halo conspiracy''~\citep{conspiracy}, and deserves attention in future studies.

\subsection{Curvature}
\label{sec:curvature}

A final significant feature of the model BTFR is its curvature, which is $\sigma_q = 3.0\: \sigma$ discrepant with the data (Fig.~\ref{fig:quad} and Table~\ref{tab:table}, final row). We caution however that this prediction is sensitive to the low-$M_\text{b}$ model uncertainties described in Section~\ref{sec:scatter}, and removing the $M_\text{b} < 10^{9.5} M_\odot$ region reduces $\sigma_q$ below $1 \: \sigma$ (Fig.~\ref{fig:quad}, green histogram).

\red{In the context of AM, BTFR curvature follows from the well-known mismatch between the slopes of the stellar and halo mass functions, and is therefore present to some extent in all AM-based studies (e.g.~\citealt{TG, Desmond}). The precise magnitude of $q$, however, depends on the details of the AM: lower curvature follows from a shallower bright end to the SMF (e.g. the photometry of~\citealt{Bernardi_SMF}, used here), a halo proxy with less concentration dependence, and a lower AM scatter. Clustering studies have begun to set strong constraints on these variables, and we find that $q$ varies by only $\sim5$ per cent as the halo proxy and AM scatter span the ranges allowed by~\citet{Lehmann}. This suggests that spatial statistics provide sufficient information on AM for the shape of the predicted BTFR to follow almost uniquely. Older AM schemes which match to halo mass directly -- as used for example in~\citet{DC} -- produce a straighter BTFR, although that change alone reduces $q$ by $15$ per cent at most and cannot lower $\sigma_q$ below $2.5 \: \sigma$. A further reduction requires a preferential decrease in halo $M_\text{vir}$ or $c$ at the faint and/or bright ends (e.g. by baryonic feedback), biases from selection effects, or systematic errors in stellar mass measurements. Some hydrodynamical simulations incorporating these effects have achieved a straighter BTFR (e.g.~\citealt{Santos-Santos, Dutton17}). $q$ likely depends in addition on the velocity measure: when sampled much beyond $R_\text{max}$, $V$ will be lowered for high-mass galaxies with falling RCs, reducing the upturn in the BTFR at the bright end. This may also contribute to the lower curvature of~\citet{DC}, who measure $V$ at $8 R_\text{d}$.}

\section{Discussion and Conclusions}
\label{sec:discussion}

\red{A range of approaches have been developed in the past two decades to elucidate the nature and origin of the baryonic Tully--Fisher relation (BTFR), but only recently have data and models become sufficiently sophisticated for statistically rigorous analysis to be possible. To advance this programme, we calculate the significance levels at which four key statistics of a state-of-the-art observational BTFR dataset, \textsc{sparc}, differ from those expected in a modern $\Lambda$CDM abundance matching model. We create mock datasets with precisely the baryonic properties of \textsc{sparc}, and analyse them in an identical fashion to the real data~\citep{Lelli}. Any differences between model and measured galaxies must therefore derive solely from differences in the distribution of dark matter, and hence be attributable to the galaxy--halo connection.} Our findings are the following:

\begin{itemize}

\item{} When defined using the flat part of the RC, the BTFR's residuals would \emph{not} be expected to anticorrelate with galaxy size in $\Lambda$CDM; a significant test of galaxy formation requires that velocity be measured at a \emph{smaller} radius, where the contribution of the baryonic mass depends more strongly on its concentration. On the other hand, the fact that the baryonic part of $V_\text{f}$ depends little on $R_\text{eff}$ makes their correlation more sensitive to a second global galaxy--halo correlation (after $M_*$--$f(M_\text{vir},c)$ imposed by AM), viz the relation between galaxy size and halo properties. We find $\sim 2 \: \sigma$ evidence for an anticorrelation of $R_\text{eff}$ with $M_\text{vir}$ or $c$ at fixed $M_\text{b}$.

\item{} The predicted BTFR scatter is $3.6\:\sigma$ larger than observed, and cannot be appreciably lowered by tightening the galaxy--halo connection. However, simulation and model uncertainties may impact the prediction at $M_\text{b} \lesssim 10^{9.5} M_\odot$, and excising this region reduces the discrepancy to $2.1 \: \sigma$.

\item{} \red{A further BTFR statistic with significant constraining power for models of the galaxy--halo connection is the quadratic curvature, which we find to be overpredicted by AM at the $3.0 \: \sigma$ level. This may indicate mass-dependent baryonic effects on the dark matter halos (e.g.~\citealt{DC1, Sawala}) or a correlation of TFR selection criteria with dynamical halo properties.}

\item{} For BTFR studies that remove galaxies with non-flat RCs (e.g. \textsc{sparc}), the fraction of such galaxies in a given dataset is orthogonal to other BTFR statistics and provides an additional handle on RC shape. We find AM to overpredict the fraction of galaxies with rising RCs at $2.2\:\sigma$, suggesting that its application to N-body halo density profiles does not fully satisfy the observed ``disc--halo conspiracy''.

\item{} The significance of discrepancies between theoretical and observed BTFR statistics is set by sample variance among model realisations, which scales inversely with the size of the data set. As statistical tests are rarely performed in the literature, the importance of sample size is often overlooked (but see~\citealt{Sorce}). Assuming BTFR statistics to obey the central limit theorem, the widths of mock data distributions -- and hence significance levels -- will vary with $\sim \sqrt{N}$, suggesting that increasing sample size may be preferable for increasing the power of statistical tests than applying stringent cuts on data quality. Because our model explicitly includes the size of the \textsc{sparc} sample (in addition to baryonic galaxy properties imported directly from the observations), our results should strictly be considered to apply only to the \textsc{sparc} BTFR, not to the BTFR per se.

\end{itemize}

We propose three directions in which this work could be taken: 1) seek the features of the BTFR -- or galaxy dynamics more generally -- with most constraining power for galaxy formation, and calculate the significance of their deviations from AM predictions; 2) modify the model to alleviate the aforementioned discrepancies, for example by complexifying the galaxy--halo connection or introducing new degrees of freedom for baryonic effects; 3) incorporate improvements in simulation resolution and survey depth to strengthen the AM prediction for $M_\text{b} < 10^{9.5} M_\odot$. This regime is not only critical for the statistical power of BTFR tests, but also connects to topical galaxy formation issues at the dwarf scale.

\section*{Acknowledgements}

I thank Federico Lelli for guidance with the \textsc{sparc} data, and Federico Lelli, Stacy McGaugh, Risa Wechsler and an anonymous referee for comments on the manuscript.

This work used the DarkSky simulations, made using an INCITE 2014 allocation on the Oak Ridge Leadership Computing Facility at Oak Ridge National Laboratory. I thank the DarkSky collaboration for creating and providing access to these simulations, and Sam Skillman and Yao-Yuan Mao for running \textsc{rockstar} and \textsc{consistent trees} on them.

This work received support from the U.S.\ Department of Energy under contract number DE-AC02-76SF00515.

\bsp


\begin{thebibliography}{10}

\bibitem[\protect\citeauthoryear{Behroozi, Wechsler \& Wu}{Behroozi et al.}{2013}]{Rockstar}
Behroozi P.~S., Wechsler R.~H., Wu H.-Y., 2013, ApJ, 762, 109 

\bibitem[Bernardi et al.(2013)]{Bernardi_SMF}
Bernardi M., Meert A., Sheth R.~K., Vikram V., Huertas-Company M., Mei F., Shankar F., 2013, MNRAS, 436, 697

\bibitem[\protect\citeauthoryear{Conroy, Wechsler \& Kravtsov}{Conroy et al.}{2006}]{Conroy}
Conroy C., Wechsler R.~H., Kravtsov A.~V., 2006, ApJ, 647, 201 

\bibitem[Desmond(2012)]{Desmond}
Desmond H., 2012, preprint (arXiv:1204.1497)

\bibitem[Desmond \& Wechsler(2015)]{DW15}
Desmond H., Wechsler R.~H., 2015, MNRAS, 454, 322

\bibitem[Desmond(2017)]{D17}
Desmond H., 2016, MNRAS, 464, 4160

\bibitem[Di Cintio et al.(2014)]{DC1}
Di Cintio A., Brook C.~B., Macci{\`o} A.~V., Stinson G.~S., Knebe A., Dutton A.~A., Wadsley J., 2014, MNRAS, 437, 415 

\bibitem[Di Cintio \& Lelli(2016)]{DC}
Di Cintio A., Lelli F., 2016, MNRAS, 456, L127 

\bibitem[Dutton(2012)]{D12}
Dutton A.~A., 2012, MNRAS, 424, 3123 

\bibitem[Dutton et al.(2017)]{Dutton17}
Dutton A.~A. et al., 2017, MNRAS, 467, 4937 

\bibitem[Kravtsov et al.(2004)]{Kravtsov}
Kravtsov A.~V., Berlind A.~A., Wechsler R.~H., Klypin A.~A., Gottlober S., Allgood B., Primack J.~R., 2004, ApJ, 609, 35 

\bibitem[Lehmann et al.(2017)]{Lehmann}
Lehmann B.~V., Mao Y.-Y., Becker M.~R., Skillman S.~W., Wechsler R.~H., 2017, ApJ, 834, 37

\bibitem[\protect\citeauthoryear{Lelli, McGaugh \& Schombert}{Lelli et al.}{2016a}]{Lelli}
Lelli F., McGaugh S.~S., Schombert J.~M., 2016, ApJ, 816, L14 

\bibitem[\protect\citeauthoryear{Lelli, McGaugh \& Schombert}{Lelli et al.}{2016b}]{SPARC}
Lelli F., McGaugh S.~S., Schombert J.~M., 2016, AJ, 152, 157

\bibitem[Santos-Santos et al.(2016)]{Santos-Santos}
Santos-Santos I.~M., Brook C.~B., Stinson G., Di Cintio A., Wadsley J., Dominquez-Tenreiro R., Gottlober S., 2016, MNRAS, 455, 476 

\bibitem[Sawala et al.(2016)]{Sawala}
Sawala T. et al., 2016, MNRAS, 457, 1931 

\bibitem[Skillman et al.(2014)]{DarkSky}
Skillman S.~W., Warren M.~S., Turk M.~J., Wechsler R.~H., Holz D.~E., Sutter P.~M., 2014, preprint (arXiv:1407.2600)

\bibitem[Sorce \& Quan(2016)]{Sorce}
Sorce J.~G., Quan G., 2016, MNRAS, 458, 2667 

\bibitem[Trujillo-Gomez et al.(2011)]{TG}
Trujillo-Gomez S., Klypin A., Primack J., Romanowsky A.~J., 2011, ApJ, 742, 16

\bibitem[van Albada \& Sancisi(1986)]{conspiracy}
van Albada T.~S., Sancisi R., 1986, Phil. Trans. Roy. Soc. Lon. Ser. A, 320, 447

\bibitem[Warren(2013)]{Warren}
Warren M.~S., 2013, Proc. Int. Conf. High Perform. Comput. Netw. Storage Anal., (New York: ACM), p. 72

\end{thebibliography}
\end{document}